\documentclass[twocolumn,twoside,slac_two]{revtex4}
\usepackage{graphicx}
\usepackage{fancyhdr}
\begin{document}

\title{Discovery of VHE Gamma-ray Emission from the Starburst Galaxy M82}

\author{Niklas Karlsson, for the VERITAS collaboration}
\altaffiliation{see http://veritas.sao.arizona.edu/conferences/authors?icrc2009}
\affiliation{Department of Astronomy, Adler Planetarium, Chicago, IL 60605, USA}

\begin{abstract}
The galaxy M82 has long been considered a promising target for very-high-energy (VHE) gamma-ray observations because of the compact starburst region in its core. Theoretical predictions have suggested it should be detectable by ground-based imaging Cherenkov telescopes like VERITAS and that a detection would have implications for the understanding of the origin of cosmic rays. M82 was observed with the VERITAS array during the 2007-2009 observing seasons. With an exposure of 137 hours, VERITAS was able to detect a gamma-ray signal at the 5$\sigma$ level. This marks the discovery of gamma rays not only from M82 but also from the new source class of starburst galaxies. The observed flux from M82 is $(3.7\pm 0.8_{\mathrm{stat}}\pm 0.7_{\mathrm{syst}})\times 10^{-13}$ photons cm$^{-2}$ s$^{-1}$ above an energy threshold of 700 GeV, which corresponds to 0.9\% of the Crab Nebula flux. The differential energy spectrum is a power law with a photon index $\Gamma=2.5\pm 0.6_{\mathrm{stat}}\pm 0.2_{\mathrm{syst}}$. Both the flux and the photon index are close to recent theoretical predictions. The VERITAS data indicate a strong correlation between the star-formation activity and the cosmic-ray production in M82.
\end{abstract}

\maketitle

\thispagestyle{fancy}

\section{Introduction}
The existence of cosmic rays near the Earth has been well established ever since their discovery in the early 20th century \citep{butt2009}, but their origin has eluded us ever since. Although there are no direct measurements of cosmic rays beyond Earth, there are indirect observations which give strong evidence for cosmic rays permeating the entire Galaxy and most significantly in the Galactic plane. This comes from the diffuse gamma-ray emission seen by the Fermi Large Area Telescope (Fermi-LAT), interpreted as mainly coming from cosmic-ray ions interacting with interstellar gas producing neutral pions which subsequently decay into gamma rays, although there is also a contribution from cosmic-ray electrons upscattering ambient photons to gamma-ray energies.

The leading hypothesis is that winds and supernovae of massive stars provide the major acceleration sites of Galactic cosmic rays, and observations of supernova remnants, such as the shell-type supernova remnant RX J1713.7-3946 \citep{aharonian2005}, are suggestive of this scenario. The detection of the Large Magellanic Cloud (LMC) with the Fermi-LAT \citep{knodelseder2009} shows gamma-ray emission coincident with the star-forming region 30 Doradus in the LMC. If massive star winds and supernova remnants accelerate cosmic rays then star-forming regions like this should indeed be emitting gamma rays.

The bright galaxy M82 is located at a distance of about 3.4 Mpc in the direction of the constellation Ursa Major \citep{sakai1999}. Ogoing interactions with nearby galaxies, including the spiral galaxy M81 (see \cite{yun1994}), have formed a compact active starburst region, about 1000 light yrs across, in the center of M82. In this region, stars are being formed at a rate approximately ten times faster than in an entire ``normal'' galaxy like the Milky Way and the rate of supernovae is 0.1-0.3 yr$^{-1}$ \citep{fenech2008, kronberg1985}. Observations of the central region of M82 at radio frequencies, interpreted as synchrotron radiation of cosmic-ray electrons spiraling in the galactic magnetic fields, suggest a very high cosmic-ray energy density, about two orders of magnitude higher than in the Milky Way \citep{rieke1980}. The mean molecular gas density is also high, about 150 particles per cubic centimeter or about $10^{9}$ solar masses total in the starburst region \citep{weiss2001}.

Because of the aforementioned properties of the starburst core, M82 has long been considered a probable gamma-ray source. Neither the Energetic Gamma-Ray Experiment Telescope (EGRET) \citep{blom1999} onboard NASA's Compton Gamma-Ray Observatory nor previous ground-based gamma-ray observatories, such as the Whipple 10m telescope \citep{nagai2005}, detected gamma-ray emission from M82. Upper limits on the gamma-ray flux were set by EGRET at 4.4 photons cm$^{-2}$ s$^{-1}$ for $E>100$ MeV and by Whipple at about 10\% of the flux from the Crab Nebula for $E>100$ GeV. The latter limit is above the sensitivity of the Very Energetic Radiation Imaging Telescope Array System (VERITAS).

\section{Observations and Analysis}
VERITAS \citep{weekes2002} is an array of four imaging atmospheric Cherenkov telescopes located at the basecamp of the Fred Lawrence Whipple Observatory near Tucson, Arizona. The array has been fully operational since mid-2007 and has the sensitivity to detect a point source with 1\% of the steady Crab nebula flux in less than 50 hours\footnote{The integral flux sensitivity above 300 GeV was improved by about 30\% with the move of one telescope in the summer of 2009, which corresponds to a decrease in the time required to detect a point source with 1\% of the steady Crab flux to about 30 hours. This did not have any impact on the M82 data set.}. Observations cover the energy range from 100 GeV to beyond 30 TeV with an energy resolution of 15-20\% above 300 GeV, and an angular resolution per gamma-ray photon of 0.1$^{\circ}$ at 1 TeV. 

The observations of M82 were made in the period between January 2008 and April 2009. The total exposure amounts to about 137 hours of quality-selected live time, i.e. time periods of astronomical darkness and clear sky conditions, which is the deepest exposure taken with VERITAS to date. By observing the source offset from the center of the field-of-view, simultaneous estimation of the background was made possible \citep{formin1994}. The source was observed at a mean zenith angle of $39^{\circ}$. 

\begin{figure}
\begin{center}
\includegraphics[width=3.2in]{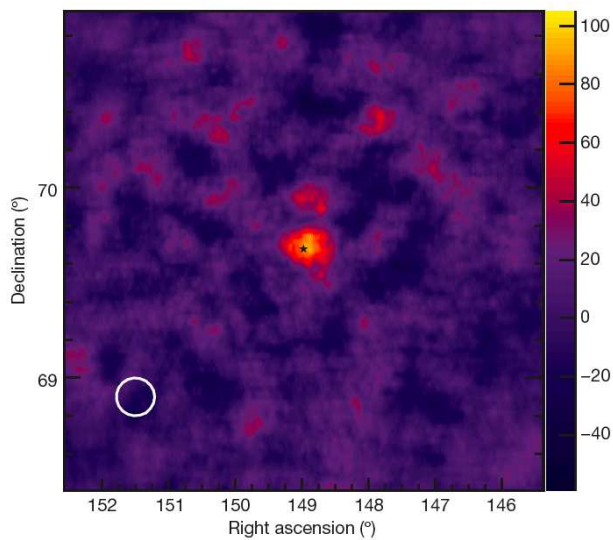}
\end{center}
\caption{Sky map showing the measured excess of VHE gamma-ray-like events above the estimated background. The excess is centered on the location of M82. Each pixel in the map contains the excess in a circular region of radius 0.1$^{\circ}$ and the background in the pixel is estimated using an annulus centered on the pixel. The spatial distribution of the excess is consistent with a point-like source. The circle is a representation of the VERITAS point spread function (68\% containment) for individual gamma rays. The black star denotes location of the core of M82. (Figure from \citep{acciari2009}.)}
\label{figure:skymap}
\end{figure}

\begin{figure}
\begin{center}
\includegraphics[width=3.2in]{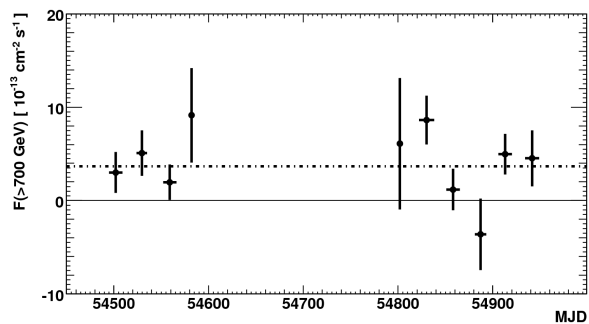}
\end{center}
\caption{Time-averaged gamma-ray flux during each monthly observation period. The integral flux was calculated with the assumption of a power-law spectrum with a photon index $\Gamma=2.5$. The error bars represent $1\sigma$ statistical errors only. The horizontal bars reflect the actual range of dates for each exposure. The dashed line shows the fit of a constant value, indicating a steady steady gamma-ray flux. (Figure from supplementary material to \cite{acciari2009}.)}
\label{figure:lightcurve}
\end{figure}

\begin{figure}
\begin{center}
\includegraphics[width=3.2in]{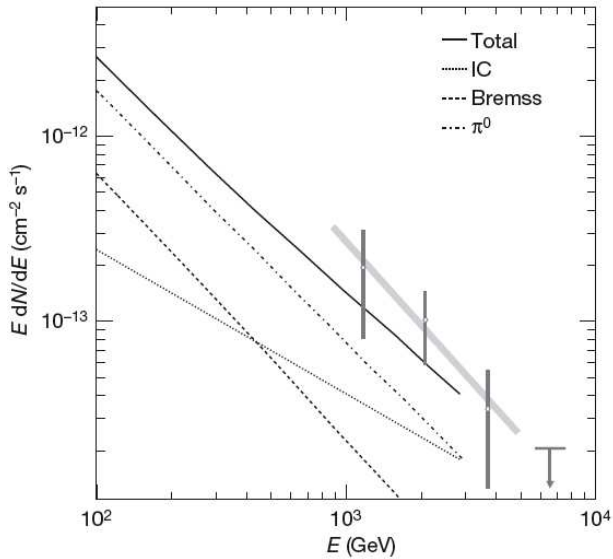}
\end{center}
\caption{Differential energy spectrum ($E\times dN/dE$, where $E$ is the gamma-ray energy and $N$ is the photon count) of M82 measured with VERITAS (open diamonds with $1\sigma$ error bars). The VERITAS points can be fitted with a power-law function, $dN/dE\propto E^{-\Gamma}$, where $\Gamma=2.5\pm 0.6_{\mathrm{stat}}\pm 0.2_{\mathrm{syst}}$, indicated by the thick grey line. The flux upper limit at about 6.6 TeV is above the extrapolation of the fitted power law at that energy. The theoretical model prediction \citep{persic2008}, of the total emission is given by the thin solid line. Its components are from $\pi^{0}$ decays, inverse Compton scatterings and bremsstrahlung. (Figure from \citep{acciari2009}.)} 
\label{figure:spectrum}
\end{figure}

The data analysis was performed using the standard VERITAS analysis tools \citep{daniel2008} using event-selection criteria that were optimized a priori for a low-flux hard-spectrum source. A total excess of 91 gamma-ray-like events (see the sky map in Figure \ref{figure:skymap}) were detected above the estimated background (267 background events). The excess corresponds to a post-trials significance of $4.8\sigma$. The observed gamma-ray flux above the 700 GeV energy threshold of the analysis is $(3.7\pm0.8_{\mathrm{stat}}\pm0.7_{\mathrm{syst}})\times 10^{-13}$ photons cm$^{-2}$ s$^{-1}$ and, as shown in the lightcurve in Figure \ref{figure:lightcurve}, there are no flux variations. From the gamma-ray flux we infer the gamma-ray luminosity to be $2\times 10^{32}$ W, which is about $2\times 10^{6}$ times smaller than the measured far-infrared (100$\mu$m) luminosity \citep{sanders2003}. The differential VHE gamma-ray spectrum (shown in Figure \ref{figure:spectrum}) is best fitted using a power-law function $dN/dE\propto E^{-\Gamma}$ with a photon index of $\Gamma=2.5\pm 0.6_{\mathrm{stat}}\pm 0.2_{\mathrm{syst}}$ (the two uncertainties corresponding to statistical and systematic errors respectively). 

The VHE gamma-ray flux from M82 is 0.9\% that observed from the Crab Nebula, which makes it one of the weakest gamma-ray sources ever detected. Because of the exceptionally long exposure, several tests were performed to ensure that systematic effects did not introduce a spurious signal in the data. The signal and the measured spectrum and flux have been verified with independent calibration and analysis chains. For further details see the supplementary information of \cite{acciari2009}.

\section{Discussion}
VERITAS has for the first time detected VHE gamma-ray emission from an extragalactic source that is not clearly assiociated with an active galactic nucleus (AGN), in which the emission of gamma rays is powered by accretion onto a supermassive black hole. It is possible that M82 harbors a supermassive black hole in its center but there is at most very weak signs of AGN activity \citep{willis1999}.

The high star formation rate in the core of M82 implies the presence of strong shocks which are expected to accelerate cosmic rays. These cosmic rays permeate the galaxy and produce gamma rays as they interact with interstellar gas and photon fields. Recent theoretical work \citep{pohl1994, pozo2009, persic2008} has predicted the VHE gamma-ray flux from M82 based on acceleration and propagation models of cosmic rays in the starburst region and the galaxy. These authors all predict fluxes close to that measured by VERITAS. In Figure \ref{figure:spectrum} the model of \cite{persic2008} is compared with the VERITAS result.

The cosmic-ray density in the starburst core of M82 can be estimated using the observed VHE gamma-ray flux. At about 250 eV cm$^{-3}$, the density is about 500 times larger than the average of the Milky Way. It must be noted though, that the Milky Way is much larger in terms of volume and thus the total energy in cosmic rays is similar.

Because of various energy-loss mechanisms, such as adiabatic cooling in the stellar winds and interactions with interstellar gas, the lifetime of cosmic rays in M82 is of the order of 1 Myr. This is lower than the typical lifetime of cosmic rays in the Milky Way by about a factor of 30. In order to maintain the same level of energy content in cosmic rays, the power input into cosmic rays must be correspondingly higher. Interestingly, the supernova rate is about 30 times higher in the starburst region of M82 than in the entire Wilky Way. The VERITAS detection of VHE gamma rays from M82 shows a correlation between cosmic-ray acceleration and massive-star formation and provides an important piece in the effort to reveal the origin of cosmic rays.

The observed VHE gamma-ray emission includes contributions from both hadronic (protons and heavier ions) and leptonic (electrons and positrons) interaction channels. The spectra from these channels are quite different in the VHE regime. Cosmic-ray ions produce gamma rays in collisions with interstellar gas. In these collisions, unstable secondary particles, mostly pions, are produced which subsequently decay. The electrically neutral pions decay directly into two photons and the charged pions decay into neutrinos and muons, which then decay into more neutrinos and electrons or positrons. The latter can then in the presence of a magnetic field emit synchrotron radiation in the radio or infrared bands.

If one assumes that the electron population is strictly secondary electrons from the above mentioned process, then an observation of the synchrotron radio emission sets an upper limit on the gamma-ray flux from cosmic-ray ions. The observations of M82 at 32 GHz frequency \citep{klein1988} set the limit at $2.5\times 10^{-9}$ photons cm$^{-2}$ s$^{-1}$ at a photon energy of 20 GeV (assuming the magnetic field is not much weaker than the current estimate of 8 nT).
This energy is below the detection capabilities of VERITAS, but an extrapolation of the VERITAS spectrum ($\Gamma=2.5$) to this energy would exceed the limit by about a factor of two. This means the spectrum at TeV energies has a slightly lower photon index than the VERITAS best-fit spectrum or the gamma-ray emission is not predominantly from cosmic-ray ions.

An alternative scenario is that the observed radio emission comes from primary electrons accelerated in the starburst region. These electrons would also interact with ambient photon field and upscatter those photons to hard X-ray and soft gamma-ray energies. By observing the non-thermal X-ray emission it is then possible to constrain the electron population. Observations place the non-thermal X-ray luminosity at 5 keV photon energy, not much higher than the observed VHE gamma-ray luminosity. The X-ray data place a lower limit on the interstellar magnetic-field strength at about a third of the current estimate of 8 nT. From this an upper limit on the absolute number of electrons with energies around 1 GeV can be derived. Theoretical predictions suggest the gamma-ray spectrum from Compton upscattering should be a power law with a photon index of about 2 in the 100 keV to 100 GeV energy range. Electrons with energies higher than that will lose energy quickly, and the acceleration efficiency is decreased at a characteristic energy. This implies a cut-off should be introduced in the spectrum and identification of such could help identify the underlying gamma-ray production mechanism.

Recently, the Fermi-LAT team has reported on the detection of high-energy gamma rays from M82 \citep{abdo2009}. Combination of data from VERITAS and Fermi-LAT will further help us understand the origin of cosmic rays.

\bigskip 
\begin{acknowledgments}
This research is supported by grants from the U.S. Department of Energy, the U.S. National Science Foundation, and the Smithsonian Institution, by NSERC in Canada, by Science Foundation Ireland, and by STFC in the U.K. We acknowledge the excellentwork of the technical support staff at the FLWO and the collaborating institutions in the construction and operation of the instrument.
\end{acknowledgments}

\bigskip 

\end{document}